\documentclass[twocolumn,10pt, aps]{revtex4-2}%
\usepackage[paperwidth=210mm,paperheight=297mm,centering,hmargin=2cm,vmargin=2.5cm]{geometry}
\usepackage{amsfonts}
\usepackage{amsmath}
\usepackage{amssymb}
\usepackage{bm}
\usepackage{graphicx}
\usepackage{color}
\usepackage{soul}
\usepackage{epsfig}%
\usepackage[dvipsnames]{xcolor} 
\definecolor{bleu_cite}{RGB}{0,0,255}
\usepackage[colorlinks=true,allcolors = black,citecolor=bleu_cite,  ]{hyperref} 
\usepackage{siunitx}

\def\fd{\textcolor{black}}
\def\fdd{\textcolor{black}}

\usepackage{stmaryrd}
\graphicspath{{/home/francois/Nextcloud/Data/Publications/2025/Super_solid/Resub-Jan-2026/figs_for_resub}}

\begin{document}

\title{Supersolid crystals of dipolar excitons in a lattice }

\author{C. Morin$^{1}$, C. Lagoin$^{1}$, T. Gupta$^{2}$, N. Reinić$^{2,3,4}$, A. Gupta$^5$, K.  Baldwin$^5$, L. Pfeiffer$^5$, G. Pupillo$^2$  and F. Dubin$^{1,\ddag}$}
\affiliation{$^1$ Université Côte d'Azur, CNRS, CRHEA,Valbonne, France }
\affiliation{$^2$Université de Strasbourg, CNRS, ISIS, Strasbourg, France}
\affiliation{$^3$Dipartimento di Fisica e Astronomia "G. Galilei" and Padua Quantum Technologies
Research Center, Università degli Studi di Padova, Padova, Italy}
\affiliation{$^4$INFN, Sezione di Padova, Padova, Italy}
\affiliation{$^5$ PRISM, Princeton Institute for the Science and Technology of Materials, Princeton University, Princeton, USA}
\affiliation{$^\ddag$: francois$\_$dubin@icloud.com}

\begin{abstract}
\textbf{In condensed-matter physics, long-range correlations introduce quantum states of matter that challenge intuition. For example, supersolids combine density order that manifests as symmetry-breaking spatial arrangement, and frictionless superfluid flow. However, supersolids have proven to only exist under very stringent conditions, with evidence limited to a few spontaneously fragmented superfluids observed in the  weakly-interacting regime. Here, we demonstrate a framework to realize crystalline supersolids in the strong interaction regime, by confining dipolar bosons in a lattice with long-range hopping. We show that dipolar excitons realize this lattice model. At fractional lattice fillings of one quarter, one third and one half we observe mesoscopic quantum crystals across around 100 sites that spontaneously break the lattice translational symmetry. At the same time, coherent long-range hopping induces off-diagonal long-range order such that the exciton solids are superfluids. Our numerical methods quantitatively confirm that supersolidity builds up in the ground-state of the lattice Hamiltonian.
}
\end{abstract}

\maketitle
\textbf{Introduction} By combining a crystalline structure and superfluidity, supersolid phases simultaneously break two distinct symmetries  — continuous translational symmetry broken by spatial order, and global phase symmetry broken at the onset of superfluidity. Following seminal predictions \cite{gross1957unified,andreev1969quantum,Thouless_1969,chester1970speculations,leggett1970can}, two general frameworks have been proposed to access supersolidity. In the first one, a weakly interacting Bose gas realizes a superfluid, with a density spontaneously modulated periodically in space \cite{Dalibard_2023,Spielman_2023}. In the second scenario, strong interactions lead to spontaneous quantum crystallization, where delocalized vacancies, or interstitials, ensure superfluidity in the solid matrix \cite{boninsegni2012colloquium,Prokofeev_2011,boninsegni2012colloquium}.

\fd{Following decades of experimental efforts, compelling evidence for supersolidity within the first framework have been reported in recent years \cite{leonard2017supersolid,li2017stripe,tanzi2019observation,bottcher2019transient,chomaz2019long,chomaz2023dipolar,trypogeorgos2025emerging,chisholm2026probing}. Supersolid signatures were found in periodic arrays of droplets or stripes, resulting from the spontaneous fragmentation of a Bose-Einstein condensate. By contrast, in the second framework supersolidity is expected in a crystalline structure with sites constituted by a single particle. This regime has long been pursued with Helium-4 \cite{Prokofeev_2011}, before research was extended to dipolar bosons confined in traps or lattices   \cite{Batrouni_2000,Goral2002,danshita2009stability,capogrosso2010quantum,Troyer_2010,Dutta2015,Baranov2012}. Notably, excitons of semiconductor bilayers are very promising   \cite{joglekar2006wigner,kurbakov2010quasiequilibrium,parish2011,conti2023chester,lozovik2019,Butov2020}, particularly  in electrostatic lattices of GaAs quantum wells where the dipolar Bose-Hubbard (dBH) Hamiltonian is accurately emulated  \cite{lagoin2022extended,lagoin2024evidence}. At fractional lattice fillings $N$, off-site dipolar repulsions between excitons thus promote quantum solids spontaneously breaking the symmetry of the lattice potential  \cite{lagoin2022extended}.}

\fd{Here, we show that in electrostatic lattices exciton solids are genuinely superfluid, and then realize mesoscopic supersolid crystals. Exciton superfluidity results from coherent long-range lattice hopping, spontaneously induced in the regime where radiative dissipation is collectively suppressed. Relying on a device where excitons carry a permanent electric dipole with an unprecedented magnitude, we report all three quantum solids predicted for the dBH model, at average lattice filling $N=$ 1/4, 1/3 and 1/2. For these, density and off-diagonal long-range orders are simultaneously revealed by spectrally resolved spatial interferometry, while superfluidity is directly manifested by quantized vortices. These observations unambiguously signal mesoscopic supersolid crystals, realized across around 100 sites of the electrostatic lattice.}

\textbf{Lattice Hamiltonian} \fd{Figure 1.a illustrates that we study dipolar excitons made by Coulomb attractions between electrons and holes confined in two distinct GaAs quantum wells (Methods).} Excitons are characterized by their permanent electric dipole \fd{(arrows in Fig.1.a)}, perpendicular to the GaAs bilayer. It enables their confinement in a gate-defined 250 nm period square electrostatic lattice. There, excitons emulate the \fdd{single-band} dipolar Bose-Hubbard Hamiltonian, \fd{in the regime where lattice sites have an occupation bound to unity, i.e. 0 or 1 (Methods)}. The dBH model is controlled by the interplay between nearest-neighbour (NN) tunneling $t$, and dipolar repulsions $V(r)$ between occupied lattice sites separated by a distance $r$ (Fig.1.a). \fd{Superfluid order then competes with theoretically up to 3 solid phases, at $N$= 1/2, 1/3 and 1/4 \cite{capogrosso2010quantum,Troyer_2010,Dutta2015,Baranov2012,lagoin2022extended,greiner_2023}.}

\fd{Along with dipolar repulsions between excitons, photon-mediated correlations provide a second channel structuring many-body ground-states in the lattice \cite{lagoin2024evidence}. Indeed, excitons are matter-wave quasi-particles that also realize optical dipoles, with radiations that extend over macroscopic distances. Exciton wavefunctions are then dressed by successive emission and re-absorption of virtual photons between distant lattice sites. As shown in Fig.1.a, long-range hopping is thus introduced in the dBH model \cite{Lehmberg,chang2018}, with coherent and incoherent contributions, $J$ and $\Gamma$ respectively.}

\fd{Figure 1.b displays the photoluminescence (PL) decay at 12 mK, for a wide range of delays $\tau$ after extinction of the laser pulse injecting carriers in the lattice. First, we observe an exponential decrease at short delays, which reveals the excitons bare radiative lifetime in the high-collisional regime ($\sim50$ ns, \fdd{see Methods}). By contrast, radiative dissipation is algebraically suppressed at lower densities, for 150 ns $\lesssim\tau\lesssim$ 450 ns (gray area in Fig.1.b). This algebraic scaling signals that long-range hopping induces interfering pathways for radiative dissipation \cite{lagoin2024evidence}. Sub-radiant many-body ground-states then emerge, with radiative decays unveiling $\Gamma$ \cite{lagoin2024evidence,chang2018}. On the other hand, the unitary counterpart of $\Gamma$, $J$, has not been evidenced yet. Theoretically, it delocalizes exciton wave-functions \cite{gold2022spatial}, as supported by a threshold increase of time coherence observed for solid ground-states \cite{lagoin2024evidence}.}

\textbf{Dipolar quantum solids} \fd{Figure 1.d} displays characteristic PL profiles in the sub-radiant regime, measured by varying the average filling $N$ through both the laser mean power density (from 600 to 800 pW.$\mu$m$^{-2}$), and the delay $\tau$ at the detection (between 150 ns and 450 ns). \fd{The left panel in Fig.1.d} shows that at $\tau=$150 ns, i.e. for $N\lesssim1$, the PL is spectrally unstructured. On the other hand, \fd{the three other panels} scrutinize lower fillings, measured by increasing $\tau$ from 250 ns to 450 ns \fd{(left to right)}. These PL spectra evidence three distinct regimes where a single narrow-band line is dominant. \fd{Recent works have shown that such structure is characteristic of spatial order  \cite{lagoin2022extended,lagoin2022mott,lagoin2023dual,lagoin2024superlattice}}.

\fd{Quantum solids predicted for the dBH model at $N=$ 1/4, 1/3 and 1/2 (Fig.1.c) have mean-field energies obtained by solely considering the strength $V$ of NN dipolar repulsions. Indeed, for a cluster made by around 10x10 sites, as in the measurements reported in Fig.1.d, we account for all possible dipolar repulsions by relying on the 1/$r^3$ scaling of $V(r)$. For our device, where we estimate that $V\sim140\mu$eV (Methods), we deduce that the mean-field energy for CB order exceeds by around 170 $\mu$eV the one of the stripe phase, and by 250 $\mu$eV the one of the star solid. The vertical lines in Fig.1.d verify that these magnitudes well account for our observations.} 

\fd{To further confirm that narrow-band profiles signal solid phases, we monitored the exciton compressibility. It is accessed by statistically evaluating the standard deviation $\sigma$ of the PL intensity \cite{lagoin2022mott}. For the three narrow band emissions discussed above, Fig.1.e highlights that $\sigma$ is  minimized only at the energy of the  dominant PL peak. It shows that a single incompressible phase is realized in each measurement, the one corresponding to the dominant PL line. This is a strong indicator that each narrow PL emission marks the CB, stripe and star solid. Experimentally, these are detected through PL profiles where the dominant solid emission accounts for around 50 $\%$ of the integrated intensity. We attribute this limitation as the result of density fluctuations, which lead to different dissipatively prepared ground-states accumulated during our minutes long stroboscopic integrations (Methods).}

\textbf{Off-diagonal long-range order} \fd{We performed Quantum Monte Carlo and tensor network calculations to compute the zero-temperature phase diagram of the lattice Hamiltonian, considering clusters around 10x10 sites (Methods). For the contribution of the dBH model, we  imposed $V$= 140 $\mu$eV and a NN tunneling rate $t$= 2 $\mu$eV \cite{lagoin2022extended}. We further included long-range hopping, restricting to its coherent part $J$ since dissipation is algebraically suppressed. Figure 2.a shows that the three solid ground-states characteristic of the dBH model are recovered, at fractional fillings $N=$ 1/4, 1/3 and 1/2 for $J/V\ll1$. This regime well corresponds to our experimental parameter space, since we estimate that $J/V$ does not exceed 0.02, as underlined by the hatched area in Fig.2.a (see Methods). On the other hand, beyond a critical $J/V$, or at fillings sufficiently different from 1/2, Fig.2.a verifies that the ground-state is superfluid (SF).}

Without long-range hopping, i.e. for the dBH model with $J=0$, extended coherence is restricted to the SF phase, and to the surroundings of the CB and star solids where extra particles or vacancies promote defect-induced super-solidity \cite{capogrosso2010quantum,Dutta2015}. When coherent long-range hopping is introduced, our exact computations confirm this behaviour \fd{(see gray area in Fig.2.a)}, but more strikingly, we observe that off-diagonal long-range order, and superfluidity, are in fact imposed in every ground-state. The three quantum solids -- CB, stripe and star -- then genuinely realize mesoscopic supersolid crystals. \fd{Figure 2.b-c illustrate this behaviour for the CB phase.} It shows that the single particle density matrix becomes constant at large distances \fd{(Fig.2.c)}, signalling off-diagonal long-range order. At the same time, the superfluid stiffness is greater than one \fd{(Fig.2.b)}, which reveals superfluidity \cite{pollock1987path}. Finally, the finite static structure factor \fd{(Fig.2.b)} attests density order and then the supersolid character of the CB phase.

To confirm experimentally that $J$ induces off-diagonal long-range order, we monitored the excitons degree of spatial coherence. It is \fd{mapped by the photoluminescence field, which carries the fingerprints of the excitons many-body wave-function responsible for its emission \cite{Anankine_2017}}. Experimentally, we use a modified Mach-Zehnder interferometer to probe the phase coherence between the PL and its image flipped along the horizontal axis (Methods). \fd{At the interferometer output, the contrast of horizontal fringes at an ordinate $y$ then quantifies the degree of phase coherence between the PL emitted at $y$ and ($-y$) (Methods).}

\fd{Figure 3} reports the PL interference at selected lattice fillings, namely $N\sim0.6$ (at $\tau=200$ ns in \textbf{a}), $N\sim0.4$ (at $\tau=400$ ns in \textbf{c}) and $N\sim0.2$ (at $\tau=500$ ns in \textbf{d}). \fd{Interference fringes are observed throughout this range,  unveiling spatially coherent exciton phases for all fillings. Furthermore, \fdd{the interference visibility, i.e. the amplitude of the interference modulation, remains large across the entire PL emission, up to 4 $\mu$m spatial separation in Fig.3.b}. This distance is four-fold larger than the thermal de Broglie wavelength at 12 mK. Excitons have then inherited off-diagonal long-range order, which is observed up to the Kelvin regime (Extended Data Fig.2). This range lies well above the limit set by quantum degeneracy in our very-dilute regime (around 30 mK \cite{Anankine_2017}). Our observations then indicate that macroscopic coherence does not result from Bose statistics, but rather from $J$ that barely varies up to the Kelvin regime. This conclusion is supported by Extended Data Fig.6, which verifies that the single-particle density matrix exhibits a decay weakly varying up to Kelvin temperatures.}

\textbf{Quantized vortices} The measurements reported in \fd{Fig.3} directly signal \fd{off-diagonal long-range order. To probe whether superfluidity is accordingly induced,} we studied the spontaneous emergence of quantized vortices that reveal the superfluid phase stiffness. Experimentally, we exploit transient electrostatic fluctuations inherent to our electrostatic confinement. The lattice indeed exhibits regions with homogeneous confinement that extend up to around 15x15 sites. Transient electrostatic defects are often observed on the edges of these domains, and typically remain static during few tens of seconds. \fd{Thereby, they provide anchoring points so that quantized vortices are stroboscopically detected \cite{Anankine_2017}.}

Figure 4 displays interferometric measurements in the regime where the CB solid is realized at $N=1/2$. On the left hand side of Fig.4.a, we note a dislocation in the interference pattern \fd{(see circle)}, with opposite-phase fringes facing each other. This situation is quantified in \fd{Fig.4.c}, which displays vertical cuts taken on the right and left hand-sides of the dislocation (see white lines in \fd{Fig.4.a}). It shows that the PL phase jumps by $\pi$, at the vertical coordinate $y\sim0.25$ $\mu$m (gray area), \fd{as expected on opposite sides of a quantized vortex}. This conclusion is directly verified in Fig.4.b, where we compute the interference pattern for a phase-coherent PL field, with one quantized vortex \fd{localized at the position of the circle}. It shows that we quantitatively reproduce our observation by solely considering the geometry of our interferometric setup. \fd{Finally, Fig.4.d displays a control measurement acquired in the same experimental conditions. Compared to Fig.4.a, the phase dislocation is suppressed (see circle), which underlines the transient nature of electrostatic defects pinning quantized vortices.}

\textbf{Supersolid order} To simultaneously detect phase and density orders, \fd{we performed spectroscopically resolved spatial interferometry. By isolating a vertical section at the center of interference patterns, we quantify the dependence of spatial coherence over the PL energy, that is over spatial order in the lattice. Let us note that these experiments}  challenge our detection sensitivity, given that our studies are realized at ultra-low densities where the PL is spectrally detected with a signal-to-noise ratio around 1.5, even for minutes long acquisitions.

Figure \fd{5}.a shows the spectrally resolved interference pattern in the SF regime. The intensity modulation along the vertical axis signals interference fringes with \fdd{a high visibility, throughout the spatial extent of the PL (Fig.5.b). Moreover,} fringes extend across the entire  500 $\mu$eV wide PL spectrum lacking spectral structure (Fig.5.c).  Thus, Fig.5.a confirms that in the SF regime extended phase coherence is established regardless spatial arrangements. Furthermore, we deduce that the superfluid fraction is large in our experiments, since it is given by the ratio betwwen the interference visibility and its amplitude at spatial auto-correlation \cite{Anankine_2017}.

Figure \fd{5}.d displays the spectrally resolved interference for the CB phase. \fd{First, we recover that the PL spectrum is dominated by a single line, at the energy for CB order (Fig.5.f). \fdd{In Fig.5.e, we further observe that the interference visibility weakly varies across the spatial extension of the PL. Off-diagonal long-range order is then robustly established.} Figure 5.d thus combines the two fingerprints of supersolids, namely density order and macroscopic phase coherence. This observation confirms that excitons condense in a single sub-radiant state for the CB phase \cite{lagoin2024evidence}. Furthermore, phase and density orders are also observed simultaneously for the stripe solid at $N=1/3$ (Extended Data Fig.3), while our signal-to-noise-ratio does not suffice to access these signatures for the star phase at $N=1/4$.}

\textbf{Conclusions} \fd{So far, supersolid crystals were theoretically conditioned on a small fraction of defects or vacancies promoting superfluidity \cite{boninsegni2012colloquium,Prokofeev_2011}. By contrast, our studies show that the dipolar Bose-Hubbard model extended by long-range coherent hopping exhibits mesoscopic defect-free supersolid ground-states. While the performance of our experiments does not allow us to distinguish yet defect-induced and defect-free supersolids, the latter are robustly found numerically up to a few hundreds of dipolar excitons. Nevertheless, we do not expect that they persist up to the thermodynamic limit, since $J$ exhibits a true long-range decay  1/$r^{1.6}$ in our lattice \cite{lagoin2024evidence}, unlike dipolar repulsions scaling like 1/$r^3$. However, $J$ has a spatial profile tuneable by suitable device engineering, so that tailored long-range hopping could offer a route towards defect-free and large scale supersolid crystals.}

\textbf{Acknowledgments}

We would like to thank the staff at CRHEA for handling support to install the dilution refrigerator, and V. Brandli and S. Chenot for their help during nano-fabrication. Also, we appreciate the time taken by M. Holzmann and N. V. Prokof'ev for discussions. Work at CRHEA was supported by the French Agency for Research (ANR-23-CE30-50022-02 (SIX)). Work at the University of Strasbourg was supported by the French Agency for Research (ANR-23-CE30-50022-02 (SIX), ANR-21-ESRE-0032 (aQCess)), the Institut Universitaire de France (IUF), and  the Horizon Europe programme HORIZON-CL4-2021-DIGITAL-EMERGING-01-30 via the project 101070144 (EuRyQa). Work at the University of Padova and University of Strasbourg has received funding from the European Union via UNIPhD programme Horizon 2020 under Marie Skłodowska-Curie grant agreement No. 101034319 and NextGenerationEU. We acknowledge the Quantum TEA libraries \cite{qtealeaves} for tensor network simulations, developed in the scope of project PASQuanS2, and the computational resources from CINECA on the Leonardo machine. Work at Princeton University was funded by the Gordon and Betty Moore Foundation through the EPiQS initiative Grant GBMF4420, and by the National Science Foundation MRSEC Grant DMR 1420541.

\textbf{Author contributions}

A.G., K.B and L.P. realised the epitaxial growth of the GaAs heterostructure while C.M. C.L. and F.D. designed and nano-fabricated the electrostatic lattice. Experimental works, data analysis, were all performed by C.M., C.L. and F.D. T.G, N.R. and G.P. realized Quantum Monte Carlo and Tensor Network computation and numerical data analysis.  F.D. designed the research project.

\textbf{Data availability}

Source data supporting all the conclusions raised in this manuscript are available for download upon request to F.D.

\textbf{Financial interest}

The authors declare no competing interest.
\vspace{1cm}

\centerline{\textbf{METHODS}}

\vspace{.3cm}

\textbf{Electrostatic lattice} \fd{Figure 1.a illustrates that we study a field-effect device embedding} two 12 nm wide GaAs quantum wells, separated by a 4 nm Al$_{.3}$Ga$_{.7}$As barrier. On the surface of the device, 220 nm above the quantum wells plane, an array of gate electrodes is deposited after electron-beam lithography. It consists of rectangles (70 nm width and 185 nm height) imprinting the lattice sites, connected by 30 nm thick wires \cite{lagoin2024evidence}. \fd{Polarizing the electrodes array, at V$_\mathrm{g}=-0.65$ V, first ensures that electrons and holes each have their minimum energy states in a different quantum well, so that the excitons ground-state is spatially indirect with a permanent electric dipole $d_X$. The latter has a magnitude  given by the distance between the two quantum well centers (16 nm). Extended Data Fig.1.b verifies that we recover this amplitude, precisely 15.3 nm, from the linear dispersion of the excitons energy with V$_\mathrm{g}$. Furthermore, the electrode geometry ensures that we imprint an  electric field sinusoidally varying in space. Its interaction with $d_X$ results in a square electrostatic potential, with 250 nm period and a depth theoretically around 160 $\mu$eV. On-site dipolar repulsions have a strength one order of magnitude larger, our device is then bound to the dipolar blockade regime where the occupation of the electrostatic lattice sites is bound to one. \fdd{Moreover, for our 160 $\mu$eV lattice depth, a single Wannier state is theoretically expected in each site.} The spatial confinement is illustrated in Extended Fig.1.a, together with the profile of the excitons  (Wannier) wave-function accessible in each electrostatic lattice site (coloured shaded areas).}\\
\vspace{.2cm}

\textbf{Effective many-body Hamiltonian} The exciton lattice is described by an effective non-Hermitian Hamiltonian incorporating both coherent and dissipative long-range hopping, $J$ and $\Gamma$ respectively, together with the standard dipolar Bose–Hubbard (dBH) terms \cite{lagoin2024evidence,chang2018,Dutta2015}. The full Hamiltonian reads
\begin{align}
    \hat{H}_{\text{eff}} = &\underbrace{-t \sum_{\langle i,j \rangle} (\hat{b}_i^\dagger \hat{b}_j + \text{h.c.}) + \sum_{i<j} \frac{V}{|r_{ij}|^3} \hat{n}_i \hat{n}_j - \sum_i \mu \hat{n}_i}_{\text{dipolar Bose-Hubbard}} \notag \\
    &+ \sum_{i \leq j} \left( J_{ij} - \mathrm{i} \frac{\Gamma_{ij}}{2} \right) (\hat{b}_i^\dagger \hat{b}_j + \text{h.c.}) \quad
\label{eq:full-hamiltonian}
\end{align}
where $\hat{b}_i^\dagger$ is the creation operator for an exciton in a site $i$, $J_{ii}=0$ while $\langle i,j \rangle$ denotes nearest-neighbour pairs and $|r_{ij}|$ is normalized by the lattice period. In Eq.\eqref{eq:full-hamiltonian}, the first three terms describe the conventional dBH model for interacting bosons on a 2D lattice, including long-range dipolar density-density interactions decaying as $1/|r_{ij}|^3$, and a chemical potential $\mu$. The last term in Eq.\eqref{eq:full-hamiltonian} represents photon-induced long-range hopping. Its coherent and incoherent parts, $J_{ij}$ and $\Gamma_{ij}$ respectively, arise from the real and imaginary parts of the dyadic Green’s function. Here, we only consider the coherent hopping $J$, since we emphasize a sub-radiant regime where radiative dissipation is collectively suppressed \cite{chang2018}. Thus, we obtain a simplified hermitian Hamiltonian
\begin{align}
    \mathcal{H} = &-t \sum_{\langle i,j \rangle} \left( \hat{b}_i^\dagger \hat{b}_j + \text{h.c.} \right) + J\sum_{i<j} j_{ij} \left( \hat{b}_i^\dagger \hat{b}_j + \text{h.c.} \right) \notag \\ &+ \sum_{i<j} \frac{V}{|r_{ij}|^3} \hat{n}_i \hat{n}_j - \sum_i \mu \hat{n}_i.
    \label{eq:numericalHamiltonian}
\end{align}
where the coefficients $j_{ij}$ are normalized such that $j_{\langle i,j \rangle} = 1$. Note that these take either positive or negative values \cite{chang2018,lagoin2024evidence}.\\

\vspace{.3cm}
\fd{\textbf{Lattice parameters}} \fd{From the theoretical profile of the lattice confinement, and considering an exciton effective mass $m_X=0.25m_0$ where $m_0$ is the free electron mass, we expect that the NN tunneling strength $t$ is around 1 to 2 $\mu$eV \cite{lagoin2022extended}. Furthermore, previous measurements, for $d_X$= 12 nm in a lattice with 220 $\mu$eV depth, have revealed that NN dipolar repulsions have a magnitude $V$ from 50 to 60 $\mu$eV \cite{lagoin2024evidence}. Hence, we first deduce that $V$ is increased to around 100 $\mu$eV, given that dipolar interactions scale with $d_X^2$. In addition, our device is operated at lower lattice depth, 160 $\mu$eV compared to 220 $\mu$eV previously \cite{lagoin2024evidence}. The spatial overlap between Wannier wave-functions from NN lattice sites is then increased (Extended Data Fig.1.a), so that $V$ is enhanced by around 30$\%$ and reaches theoretically 130 to 150 $\mu$eV.}

\fd{In Eq.(2), $J$ describes the coherent contribution of photon-mediated hopping between NN sites. While $J$ is usually obtained by assuming point-like optical dipoles \cite{chang2018}, excitons are instead characterized by Wannier wave-functions in the lattice (Extended Data Fig.1.a). $J$ is accordingly enhanced by the overlap between NN sites (Extended Data Fig.1.a). It is in fact bound to the excitons self-interaction energy, of the order of the energy separation between optically dark and bright states. Then, $J$ can not not exceed a few micro-eV for our double GaAs quantum wells \cite{combescot2022signature}. Our experiments are then restricted to the region of the phase diagram where $J/V\lesssim$ 0.02 (hatched region in Fig.2.a).}

\vspace{.3cm}
\textbf{Experimental techniques} Our experiments rely on a 12 mK micro-PL setup. The lattice device is placed on 3-axis piezo-electric actuators (Onnes Technologies), mounted inside a dilution refrigerator (Bluefors LD). We rely on a 0.9 aperture microscope objective to laser excite the sample, and collect the resulting photoluminescence (PL). Precisely, we inject dipolar excitons with a laser excitation at resonance with the direct excitons absorption of the two GaAs quantum wells (at around 795 nm). The PL radiated by dipolar excitons, at around 810 nm, is monitored at controlled delays to the termination of the laser illumination, in 20 to 100 ns long intervals using an intensified time-gated CCD camera (PI-MAX from Princeton Instruments -- \fd{see Fig.1.b}). Our stroboscopic measurements are performed at a repetition rate of 700 kHz \fd{as in previous works \cite{lagoin2024evidence}}. Accumulation times then typically reach minutes, since the PL has a spectrally integrated intensity reduced to a few hundreds counts per minute. For spectroscopic measurements (e.g. in Fig.1), a 500 mm focal-length imaging spectrometer with a 1800 lines/mm grating is used.

\vspace{.3cm}
\textbf{Sub-radiant dynamics} Figure 1.b illustrates the PL dynamics in the lattice. At $\tau\lesssim$ 150 ns, the average density is large so that a high rate of dipolar collisions prevents the buildup of collective radiance. This allows us to extract the excitons bare radiative decay rate $1/\Gamma_0\sim50$ ns from the exponential decrease of the PL integrated intensity \cite{lagoin2024evidence}, in reasonnable agreement with theoretical expectations. On the other hand, Fig.1.b highlights that the PL decay is either algebraic at 150 ns $\lesssim\tau\lesssim$ 450 ns, or exponential with a time constant about 350 ns at $\tau\gtrsim$ 450 ns. In both cases, radiative dissipation is strongly suppressed, the power-law decay providing a direct signature of many-body sub-radiance \cite{lagoin2024evidence,chang2018}.

\fdd{Extended Data Fig.1 verifies that the lattice filling is bound to unity in the sub-radiant regime. For that, panel d reports the dynamics of the PL blueshift, i.e. the difference between the PL energy and its value in the most dilute regime at longest delays. This difference quantifies the exciton density \cite{ivanov2010}. In Extended Data Fig.1.d, one notes that the PL energy does not vary for $\tau\gtrsim200$ ns, within our instrumental resolution (100 $\mu$eV). The lack of measurable blueshift then sets an upper bound to the exciton concentration, precisely 2$\cdot10^9$ cm$^{-2}$ at $\tau\sim200$ ns, which lies below unitary filling.}

\vspace{.3cm}
\textbf{Mean-field energies of exciton solids} In the lattice, the dipolar potential $V(r)$ scales as $1/r_{i,j}^{3}$, where $r_{ij}$ denotes the distance between sites $i$ and $j$. The mean-field energies of solid phases are then directly expressed as a function of $V$ in the mesoscopic regime. For 10x10 sites we find that the mean dipolar interaction energy is around $2.7V$ for the CB, $1.5V$ for the stripe phase, and $0.9V$ for the star one. As mentioned above, we expect that $V$ lies between 130 to 150 $\mu$eV for our device. Critically, we quantitatively match the energy positions of the CB, star and stripe solid phases in \fd{Fig.1.d} by setting $V=140$ $\mu$eV.

\vspace{.3cm}
\textbf{Spatial interferometry} Interference measurements rely on a modified Mach-Zehnder interferometer, where one arm flips the PL image along the horizontal axis (with a retroreflector), the two output PL fields being finally recombined with a vanishing optical delay. Thus, the interference visibility quantifies the first-order spatial-correlation function, $g^{(1)}$, along the vertical direction $y$, precisely $|g^{(1)}(2y)|$. Note that we impose a vertical tilt angle between the 2 interfering fields, so that interference fringes are horizontally aligned, with a period set to around 500 nm (see Fig.4.e). \fdd{The interference contrast, measured by computing the spatially resolved amplitude of the interference modulation, is therefore limited to about 40$\%$ at auto-correlation by the reduced mode matching between interfering fields.}

\fdd{Extended Data Fig.4 verifies that in Fig.3-5 the PL spatial coherence lies well beyond the classical limit set by our optical resolution. For that, we report the interference pattern measured for a spatially incoherent light emission, obtained by reflecting a led on the surface of our device. One notes that only a single fringe is observed on each side of the spatial auto-correlation ($y=0$). In fact the interference contrast vanishes for $|y|\gtrsim1$ $\mu$m (right panel) thus quantifying our optical resolution (gray area). The latter is systematically displayed in Figs.3 to 5.}

\vspace{.3cm}
\textbf{Analysis of interference profiles} The plain lines associated to the profiles in Fig.3-5 are obtained by filtering interferometric images in Fourier space. Only the modes at zero frequency - associated to the non-modulated signal - and the dominant non-zero frequency (and its conjugate) - associated to the interference intensity modulation - are conserved by setting the contributions of other frequencies to zero. The size of the non-vanishing frequency windows around the three chosen modes is set such that the structure of the filtered signal accurately reproduces the measurements. From the ratio between the absolute value of the signal associated to the dominant non-zero frequency, and the signal associated to the zero frequency, we deduce the interference contrasts  quoted in the main text.

\vspace{.3cm}
\fd{\textbf{Purity of dissipative preparations}} \fd{Our studies are bound to the regime where $J/V\lesssim$ 0.02 and $N<1$. Figure 2.a shows that accessible ground-states then reduce to either SF, CB, stripe or star solids, with no filling where phases coexist. Each solid ground-state is signalled by a characteristic PL spectral profile, made of a single narrow-band PL line. Experimentally, this structure is detected with a finite purity, fixed by fluctuations of the average filling $N$ during each acquisition, in a 100 ns long time interval in Fig.1.d. This limitation is illustrated in Extended Data Fig.1.c, which shows two PL spectra successively measured for the same experimental conditions. These differ and in fact have spectral structures corresponding to dominant CB and stripe orders respectively. Thus, we deduce that $N$ varies by up to 30$\%$ between successive measurements. We attribute this limitation to fluctuations of both the magnitude of $N$ at the onset of the sub-radiant regime, and to variations of the algebraic decay of the excitons radiative dynamics. As a result, PL profiles average different ground-states realized while $N$ fluctuates.} 

Throughout the manuscript, PL spectra are modelled by adjusting the fraction of realizations implementing each accessible ground-state. Hence, we deduce the ratio between the integrated intensity of the individual narrow-band line of a given target solid phase, with the experimentally measured integrated intensity. This quotient defines the average purity of the dissipative preparation. It is around 50$\%$ for the solid phases discussed  throughout the manuscript.

\textbf{Numerical simulations} To explore the many-body ground-state of the Hamiltonian given in Eq.(2), we employ numerically exact large-scale quantum Monte Carlo (QMC) simulations based on worm algorithm \cite{prokof1998exact}, along with variational tensor network techniques. The worm algorithm operates in continuous imaginary time and samples configurations that contribute to both diagonal and off-diagonal observables, by extending the configuration space to include open worldline segments (``worms"). This approach is particularly effective for bosonic systems. It yields direct access to the single-particle density matrix $G$ (Fig.2.c), enabling precise characterization of off-diagonal long-range order. Additionally, the algorithm efficiently measures topological winding numbers, which provide an estimator for superfluid stiffness $Y_s$ (Fig.2.b). Complementary tensor network simulations based on a binary tree tensor network (TTN) \cite{shi2006ttn, gerster2014ttn} are used to variationally approximate the ground state and benchmark the QMC results. These simulations employ a mapping of the two-dimensional lattice model onto an equivalent one-dimensional one \cite{cataldi2021hilbert}, and optimize the TTN tensors sequentially with a fixed bond dimension.

\textbf{Validation} The presence of both positive and negative hopping amplitudes $j_{ij}$ in Eq.~\eqref{eq:numericalHamiltonian} gives rise to a sign problem in QMC, as the path-integral weights associated with different configurations are no longer strictly positive, thereby losing their probabilistic interpretation. In contrast, tensor network methods do not suffer from such a sign problem. To assess the impact of positive $j_{ij}$ couplings on the ground state, we compare the ground state properties under two scenarios: (i) using the full set of $j_{ij}$ terms (including both positive and negative couplings), and (ii) retaining only the negative $j_{ij}$ terms. 

As shown in Extended Data Fig.5.b, the removal of positive couplings does not qualitatively alter the ground state. Heuristically, while positive hopping imposes an energy penalty between certain pairs of sites, superfluidity persists via paths involving only the negative $j_{ij}$ couplings. Extended Data Fig.5.a shows the particle density $\rho$ as a function of chemical potential $\mu/V_{NN}$ at fixed $J/V_{NN}=0.02$, from both TN and QMC simulations. The plateaus at $\rho = 1/4$ and $\rho = 1/2$ indicate the star and checkerboard supersolids, respectively, and their persistence across both models confirms that the phase diagram remains unchanged. The striped phase at $\rho = 1/3$, on the other hand, is not observed on the $8 \times 8$ lattice, as the total number of sites has to be divisible by three in order to detect stripe order for finite-sized systems. Extended Data Fig.5.b presents the single-particle density matrix, $|G(r)|=|\langle \hat{b}_i^\dagger \hat{b}_{j} \rangle|$, as a function of the site separation $r\equiv r_{ij}$, at $\mu/V_{NN} = 3.5$, i.e. inside the checkerboard region, for $J/V_{NN}=0.02$. It confirms that off-diagonal long-range order is preserved in both models. Importantly, the close agreement between TTN and QMC across both panels directly supports the robustness of observed phases.

\textbf{Observations} To further characterize the solid and superfluid components of the phases at 1/3 and 1/4 filling, we analyze both the static structure factor $S(\mathbf{k})$ and single-particle density matrix $|G(r)|$. Extended Data Fig.6.a-b shows $S(\mathbf{k})$ computed over the first Brillouin zone for two representative phases, at a fixed hopping ratio $J/V_{NN} = 0.02$. For visual clarity, the value at $\mathbf{k} = 0$, which reflects the total particle number, has been set to zero to enhance the contrast of nontrivial features. In Extended Data Fig.6.a, corresponding to a $9 \times 9$ lattice with chemical potential $\mu/V_{NN} = 2.0$, the structure factor exhibits dominant peaks at $\mathbf{k} = \pm( 2\pi/3, 2\pi/3)$, indicative of a stripe-like density modulation. Extended Data Fig.6.b shows results for a $10 \times 10$ lattice with $\mu/V = 1.4$, where peaks appear near $\mathbf{k} = (0, \pm\pi)$ and $(\pm\pi, 0)$, characteristic of a star-like density pattern. These distinct momentum-space signatures reflect the different crystalline orders of the two phases. To further probe their superfluid coherence, we examine the real-space single-particle density matrix $|G(r)| = |\langle \hat{b}_i^\dagger \hat{b}_{j} \rangle|$, shown in Extended Data Fig.6.c, as a function of the separation $r=|\textbf{r}_i-\textbf{r}_j|/a$ where $a$ denotes the lattice period. Both curves exhibit off-diagonal long-range order characteristic of a superfluid. Taken together, these results confirm the coexistence of both diagonal and off-diagonal long-range order, hallmarks of supersolids.

\clearpage
\newpage

\onecolumngrid

\centerline{\includegraphics[width=\linewidth]{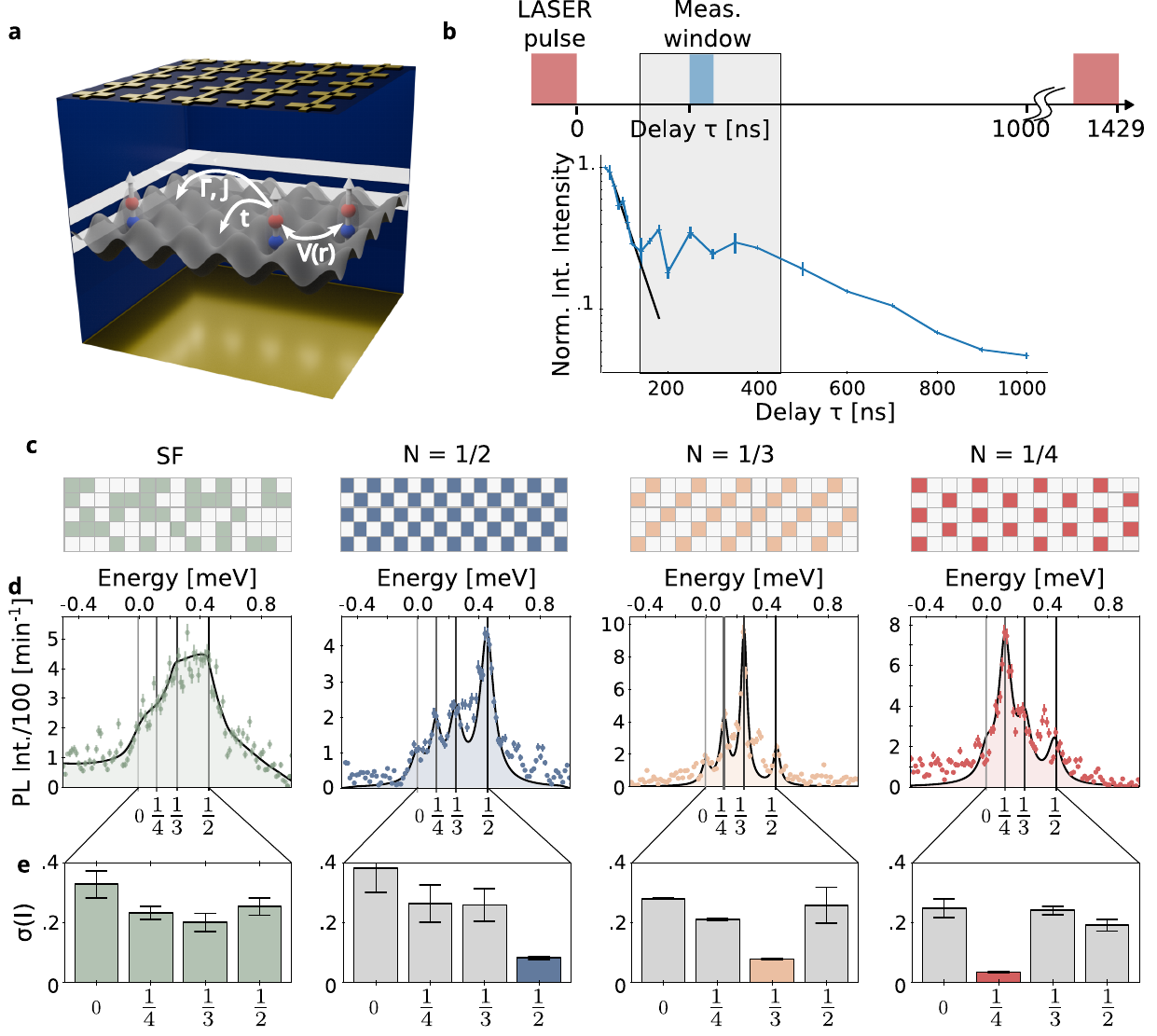}}
\fd{\textbf{Fig.1: Sub-radiant dipolar quantum crystals.} \textbf{a} Surface gate electrodes (gold) imprint an electrostatic lattice where dipolar excitons emulate the dBH model, controlled by dipolar repulsions $V(r)$, NN tunneling $t$ and long-range hopping $(J,\Gamma)$. \textbf{b} PL dynamics following laser injection of electronic carriers.  The PL decays exponentially with a time constant $1/\Gamma_0$ = 50 ns (black line) for $\tau\lesssim150$ ns, while for $\tau\gtrsim$ 150 ns radiative dissipation is suppressed by the spontaneous buildup of many-body sub-radiance. The lattice Hamiltonian is probed in this latter regime (gray area) where $N$ is best controlled. \textbf{c} Spatial arrangements corresponding to a SF phase in the lattice (left), and to the CB, stripe and star orders, at $N$= 1/2, 1/3 and 1/4 respectively. \textbf{d} PL spectra measured across 20 sites, in the SF regime (green), for CB (blue), stripe (beige) and star (red) orders. Vertical gray lines mark the energies theoretically expected for each ordered phase, as well as the one free from dipolar repulsions (0 at the lowest energy). Black curves provide model spectra obtained by adjusting the fraction of realizations implementing each accessible ground-state  (Methods). \textbf{e} Fluctuations of the PL intensity $\sigma(I)$ computed by statistically analyzing ensembles of 10 measurements. Fluctuations are spectrally resolved at the energies specified by the 4 vertical lines. Experiments have all been carried out at 12 mK, error bars displaying the poissonian precision, possibly comparable to the size of data points.}

\newpage

\includegraphics[width=\linewidth]{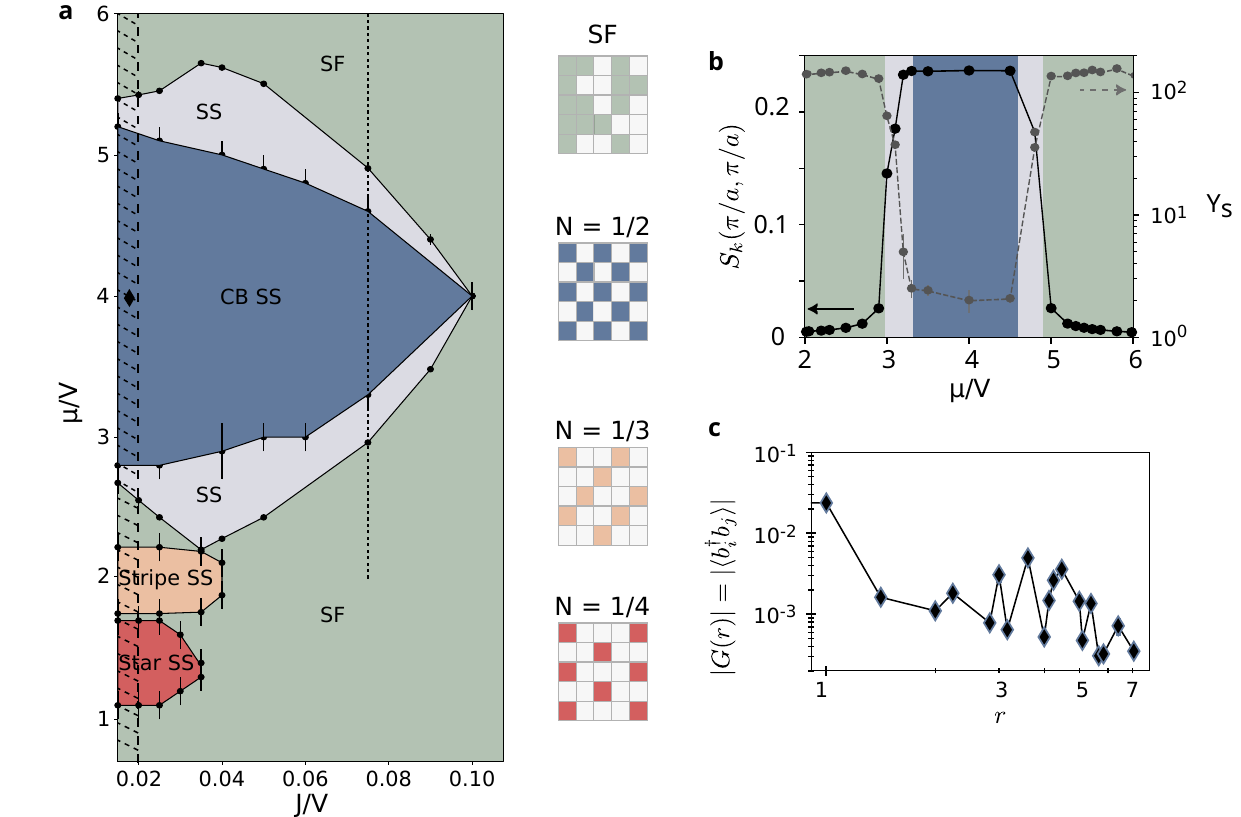}\vspace{.2cm}
\fd{\textbf{Fig.2: Phase diagram.} \textbf{a} Quantum Monte Carlo and tensor network simulations, performed on a 10x10 sites cluster, showing the phase boundaries in the ($J/V$, $\mu/V$) plane. We identify a superfluid (SF) domain (green), and different supersolid (SS) regimes: a CB SS at $N=1/2$ (blue), stripe SS at $N=1/3$ (beige), star SS at $N=1/4$ (red), and defect-induced SS (gray). The stripe SS is obtained on a 9x9 lattice, while the right panels recall the corresponding spatial arrangements. The hatched rectangle underlines the experimentally studied parameter space. \textbf{b} Static structure factor at the corner of the first Brillouin zone, $S_k(\pi/a, \pi/a)$ (in black -- left axis), and superfluid stiffness, $Y_s$ (in gray -- right axis in log scale), as a function of chemical potential $\mu/V$ for $J/V = 0.075$ (see dotted line in \textbf{a}). \textbf{c} Single-particle Green function for the CB SS, $|G(r)|$, as function of the distance $r=|\textbf{r}_i-\textbf{r}_j|/a$, computed for our experiments ($J/V=0.015$ and $\mu/V=$ 4 -- see diamond in \textbf{a}).}

\newpage

\centerline{\includegraphics[width=\linewidth]{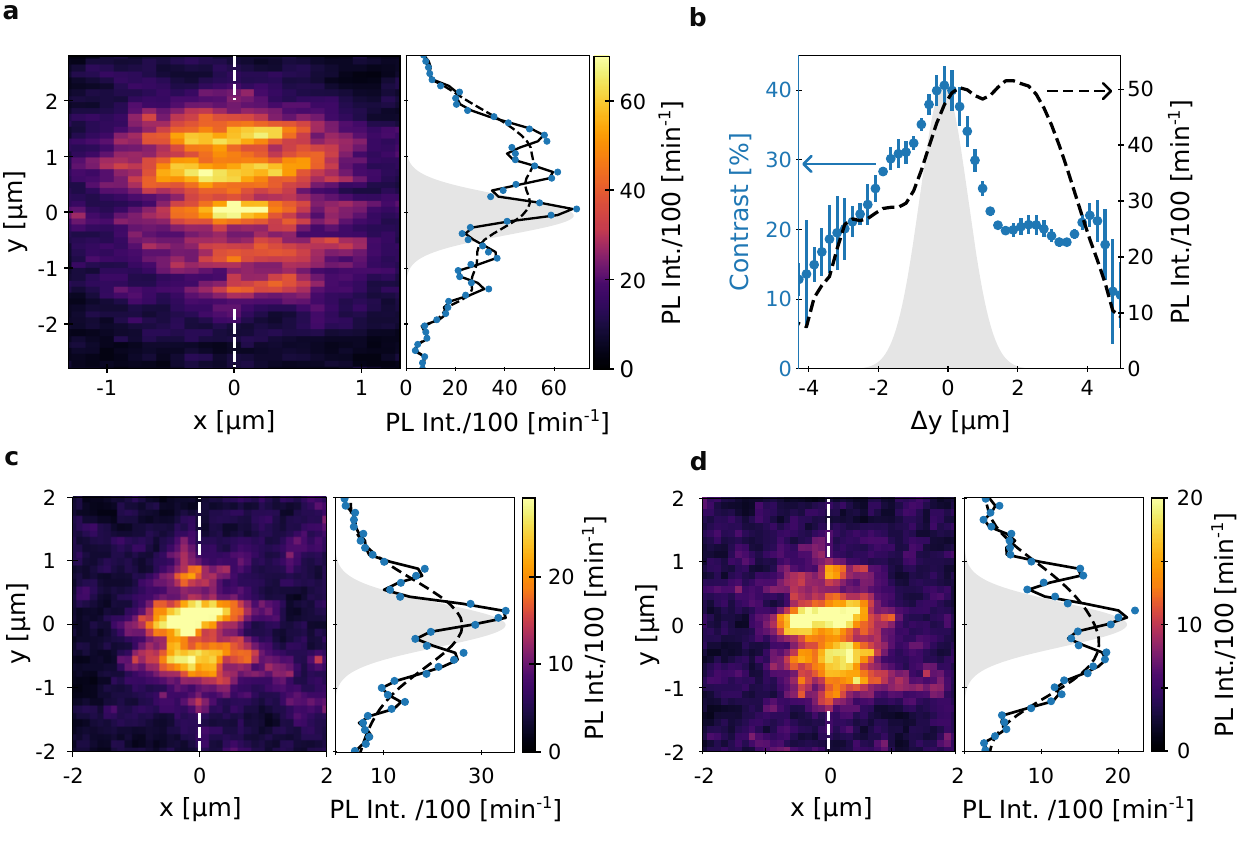}}\vspace{.2cm}
\textbf{Fig.3: Spatial coherence in the lattice.} \fdd{\textbf{a} Interference of the PL emitted when $N\sim0.6$ (at $\tau=250$ ns). \textbf{b} Interference contrast (blue) and non-interfering PL intensity (dashed line) measured along the dashed white line in panel \textbf{a}. The horizontal axis marks the spatial separation $\Delta y=2y$ between interfering regions. \textbf{c-d} Interference of the PL emitted when $N\sim0.4$ ($\tau=400$ ns \textbf{c}) and $N\sim0.2$ ($\tau=500$ ns \textbf{d}). Measurements in (\textbf{a}-\textbf{c}-\textbf{d}) are integrated in a 100 ns long time interval. Right panels provide vertical cuts of the interferograms along the white dashed lines, reproduced by filtering the interference signal. It yields a visibility around 20-30 $\%$ (solid lines) for the interference modulation throughout the emission, dashed lines providing the profile without any interference. Gray areas display our optical resolution. Experiments have all been performed at 12 mK, error bars providing poissonian precision, possibly below the size of data points.}

\newpage

\includegraphics[width=\linewidth]{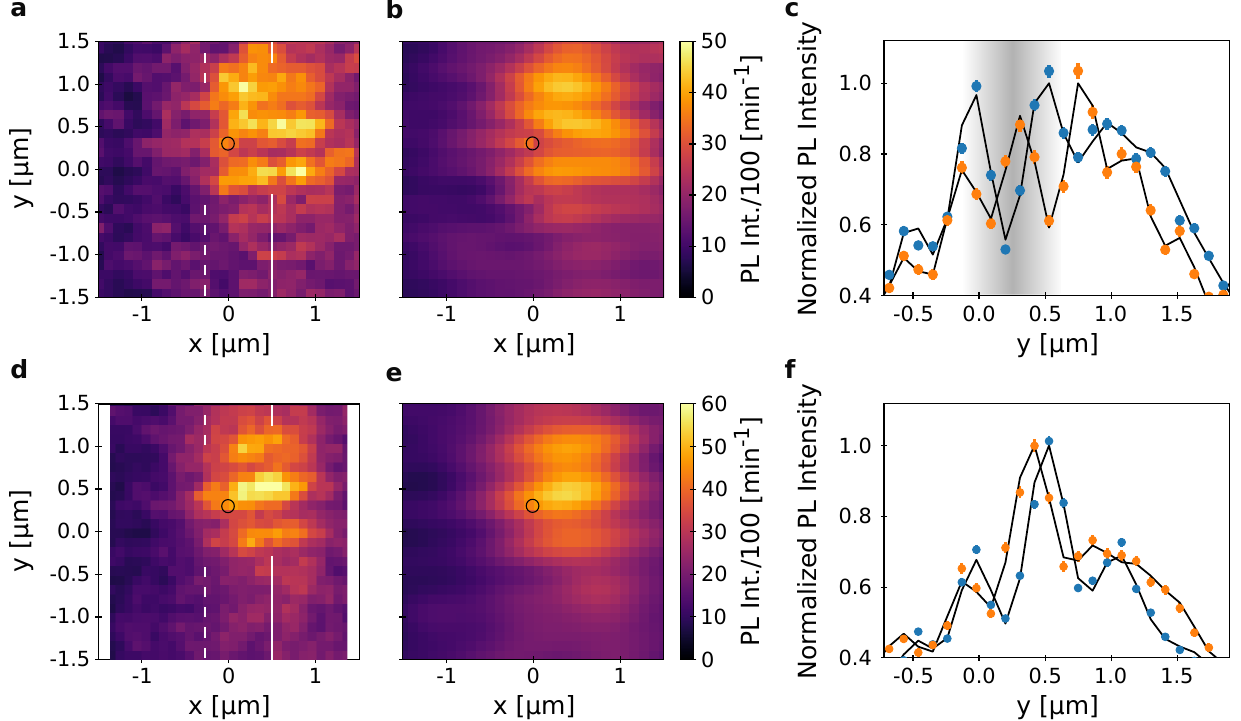}\vspace{.2cm}
\textbf{Fig.4: Superfluid stiffness.} \textbf{a} Interference pattern measured in the regime where a CB solid is dissipatively prepared. \textbf{b} Model interference profile simulated for a PL emission with a phase uniform spatially, and one quantized vortex \fd{located at the position of the circle}. \textbf{c} Vertical line cuts taken on the left (orange) and on the right (blue) hand-sides of the vortex (see dashed and solid lines in \textbf{a}). Solid curves present the variations obtained by filtering the interference signal yielding 30$\%$ visibility. \fd{\textbf{d} Interference pattern measured in the same experimental conditions as in \textbf{a}, together with the simulated pattern free from quantized vortex (\textbf{e}). \textbf{f} Line cuts taken at the positions of the solid and dashed lines in \textbf{d}, orange and blue respectively. In \textbf{c}-\textbf{f}} error bars display poissonian precision, around the points size,
while experiments were performed at 12 mK.\\
\newpage

\centerline{\includegraphics[width=\linewidth]{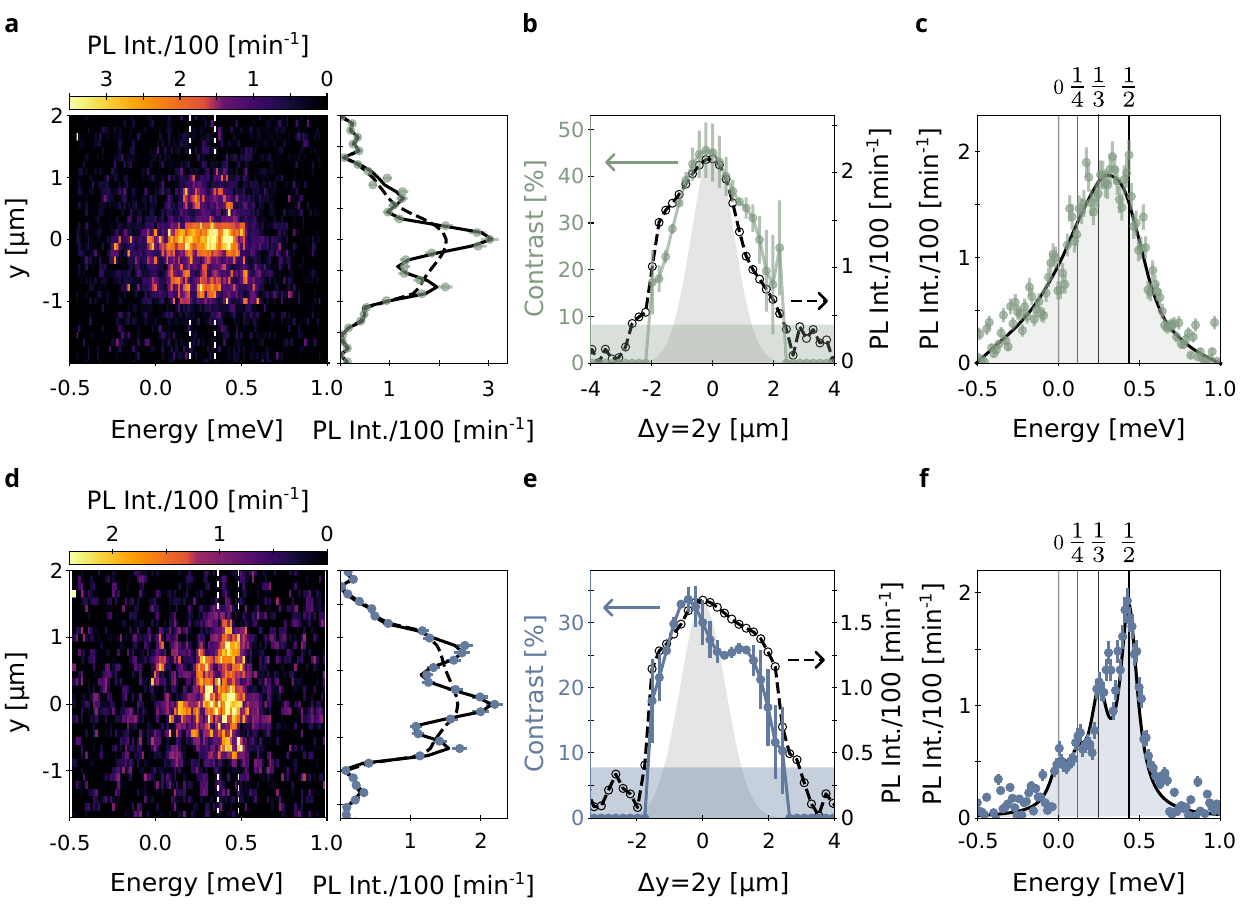}}\vspace{.2cm}
\fd{\textbf{Fig.5: Supersolid order.} (\textbf{a}) Spectrally (horizontal axis) and spatially (vertical axis) resolved interference, in the SF regime. \fdd{The right panel displays the interfering (green points -- solid line) and non-interfering (dashed black line) signals, obtained by averaging the image between the white dashed lines. \textbf{b} Interference contrast (green) across the PL emission (open circles), obtained by averaging panel \textbf{a} between the dashed lines.} \textbf{c} Spectral profile of the PL obtained by averaging along the vertical axis in \textbf{a}. The solid line provides a model spectrum. \textbf{d} Spectrally resolved interference pattern for the CB phase. \fdd{The right panel shows the interference modulation (blue -- solid line) at the energy for CB order (between white dashed lines), together with the non-interfering profile (dashed black line). \textbf{e} Interference contrast (blue) and non-interfering PL emission (open circles and dashed black line), obtained by averaging panel \textbf{d} between the white dashed lines.} \textbf{f} PL spectrum obtained by averaging over the vertical axis of panel \textbf{d}, together with a model profile (solid line). \fdd{In (\textbf{b}, \textbf{e}) the green and blue areas mark the threshold below which the PL intensity is too low to reliably extract an interference visibility, while the gray area displays our spatial resolution.} Experiments have all been performed at 12 mK, error bars displaying poissonian precision.}

\newpage

\centerline{\includegraphics[width=\linewidth]{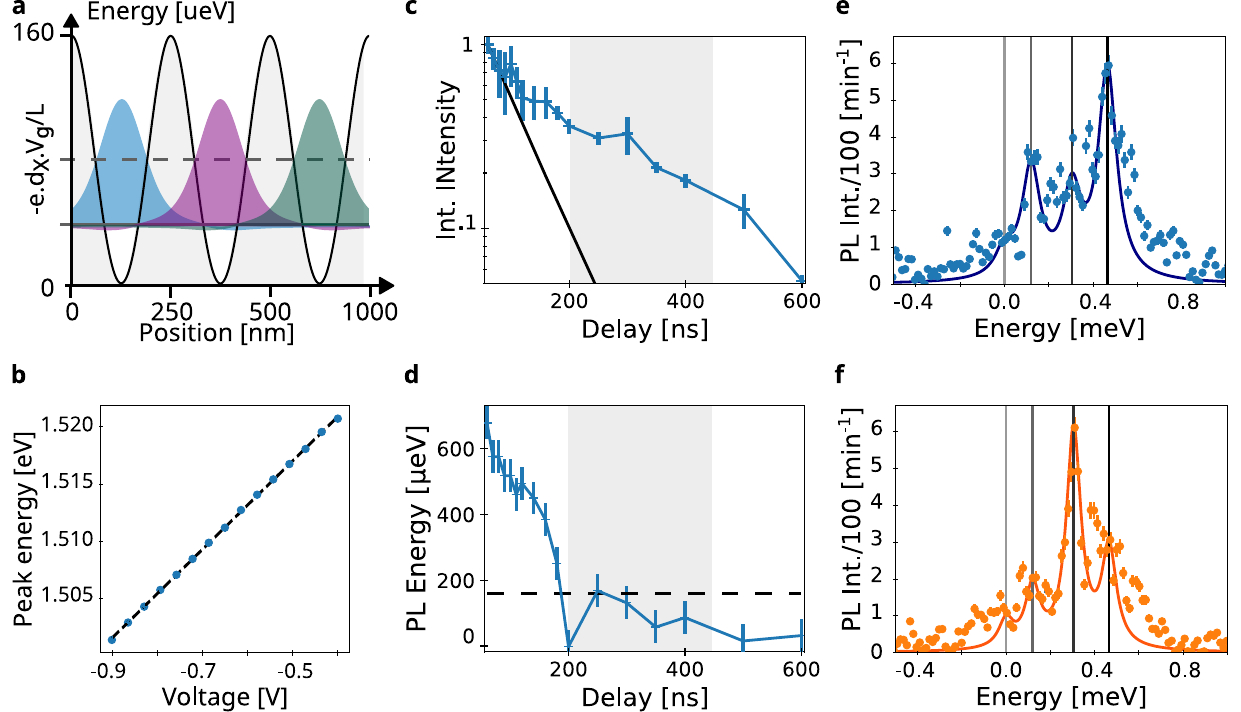}}\vspace{.2cm}
\fd{\textbf{Extended Data Fig.1: Exciton lattice Hamiltonian.} \textbf{a} Calculated profile of \fdd{the single} Wannier orbital (shaded areas) accessible in each site of our 160 $\mu$eV deep lattice. The Wannier energy is given by $-ed_X.V_g/L$, $L=$ 393 nm being the thickness of our device and $e$ the electron charge. \textbf{b} Variation of the Wannier energy as a function of the bias $V_g$ applied on top electrodes. From the linear dispersion (dashed line) we deduce $d_X=$ e$\cdot$15.3 nm. \fdd{\textbf{c-d} PL integrated intensity (\textbf{c}) and PL energy shift compared to the most dilute regime at longest delays (\textbf{d}) . In \textbf{c} the solid line marks the 60 ns excitons bare radiative decay, while in \textbf{d} the PL energy does not vary for $\tau$ beyond 200 ns. The exciton density is then bound to 2$\cdot10^9$ cm$^{-2}$ at $\tau$=200 ns, corresponding to at most one exciton per site. In \textbf{d} the dashed line marks the lattice depth.} \textbf{e-f} PL spectra for two successive measurements performed in the same experimental conditions targeting $N=$ 1/2. The profile in blue (\textbf{c}) reveals that CB order is dominant, as expected, while the orange profile (\textbf{d}) instead unveils dominant stripe order. This difference marks that $N$ is dissipatively prepared with up to 30 $\%$ fluctuations. Vertical lines mark the energies of the different solid phases, as in the main text, while solid lines provide model spectra obtained by adjusting the fraction of realizations of each accessible solid ground-state. Measurements have all been realized at 12 mK, error bars displaying poissonian precision.}

\newpage

\includegraphics[width=\linewidth]{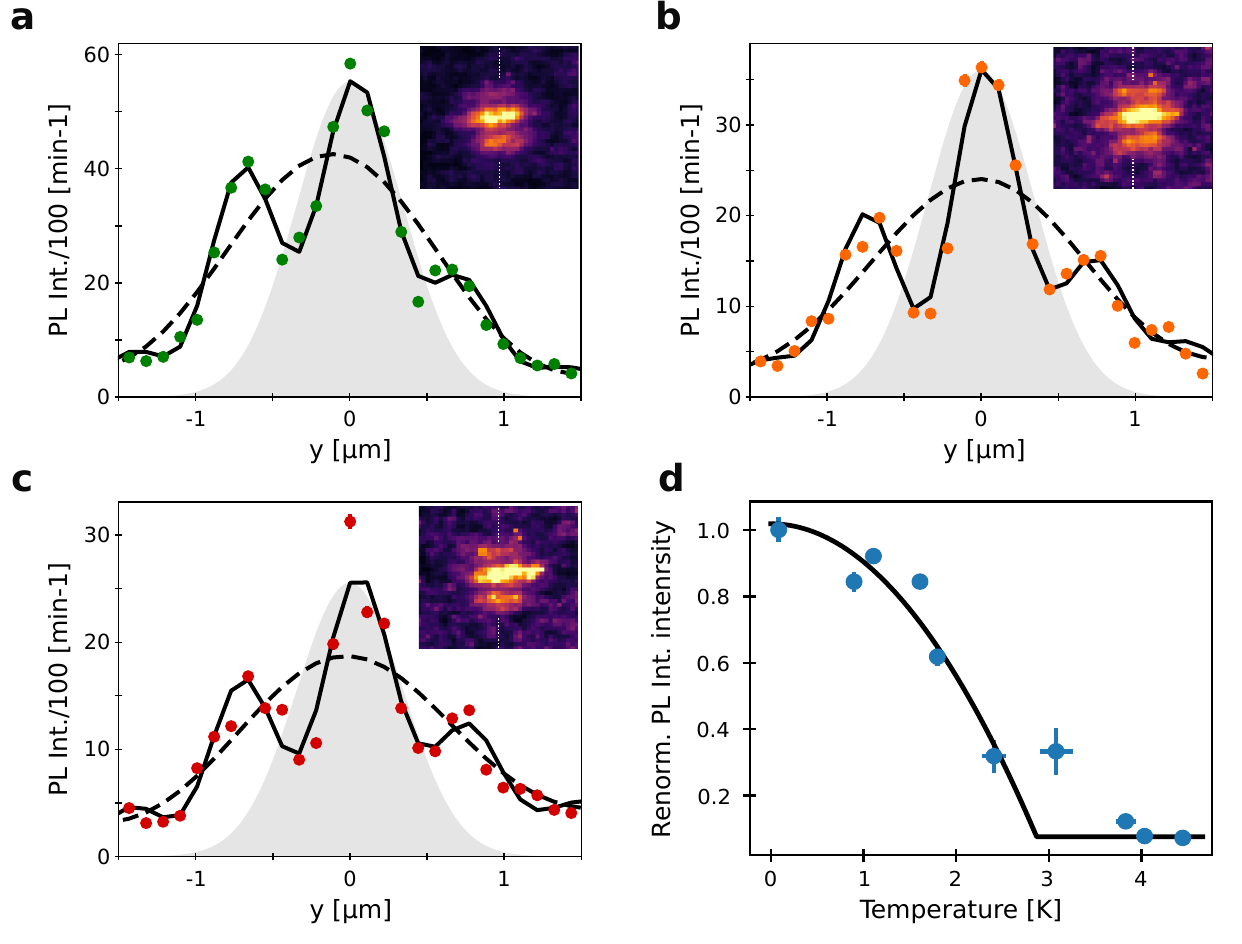}\vspace{.2cm}
\textbf{Extended Data Fig.2: Phase-coherence vs. temperature.} \textbf{a}-\textbf{c} Spatially resolved interference profiles measured at $T$=400 mK (\textbf{a}), 1000 mK (\textbf{b}) and 1300 mK (\textbf{c}). Cuts of the interference profiles are taken along the white lines displayed in each image. \fd{We deduce that the visibility lies around 30\% by filtering  the  interference signal, as shown by the solid lines, while dashed lines provide the bare spatial profile without any interference. Gray areas quantify our optical resolution. Poissonian precision lies below the size of data points. \textbf{d} Integrated photo-luminescence over the full emission profile for temperatures ranging from 12 mK to 4.5 K. Data points are acquired at a fixed delay $\tau$= 200 ns, in a 100 ns long time interval where we only detect sub-radiant phases. The solid line provides a quadratic adjustment to the data, yielding a critical temperature around 2.7 K for dissipative long-range hopping.}

\newpage

\centerline{\includegraphics[width=0.8\linewidth]{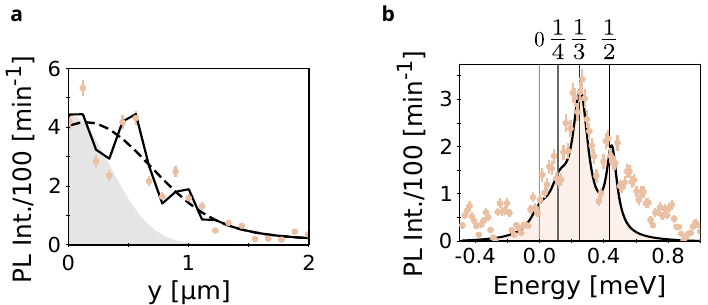}}\vspace{.2cm}
\fd{\textbf{Extended Data Fig.3: Signatures of a stripe supersolid.} \textbf{a} Variation of the PL intensity measured in spectrally resolved interference experiments (beige), together with the filtered modulation (solid line) and the bare intensity without any spatial interference (dashed line). \textbf{b} Simultaneously measured PL spectrum, together with a model profile obtained by adjusting the average fraction of realizations yielding each accessible solid ground-state (solid line). Experiments have been performed at 12 mK, error bars displaying poissonian precision.}

\vspace{.2cm}

\centerline{\includegraphics[width=\linewidth]{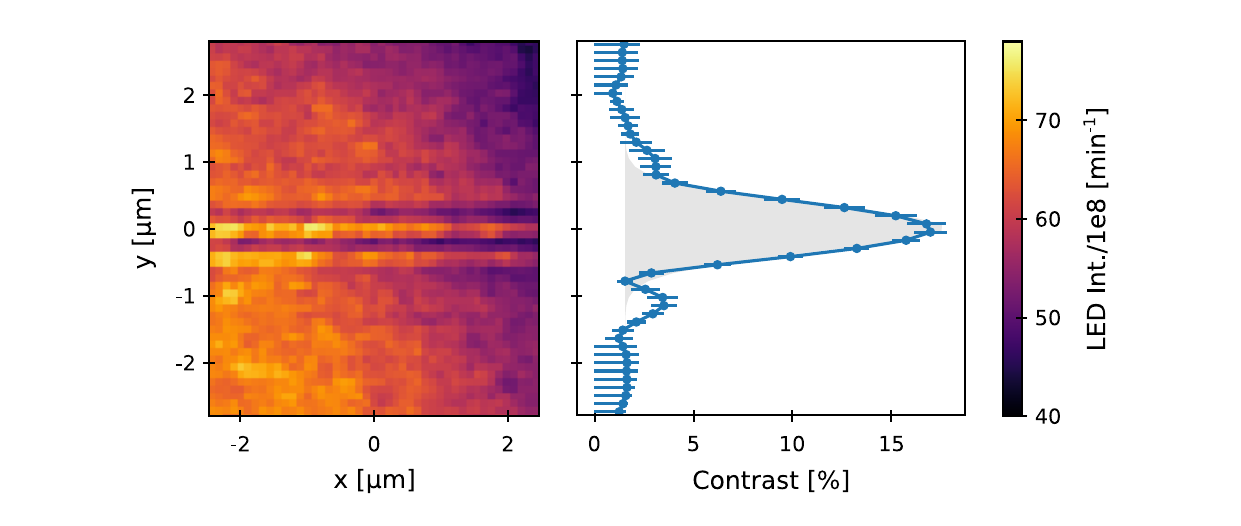}}
\vspace{.1cm}
\fdd{\textbf{Extended Data Fig.4: Interferometry spatial resolution.} Interference profile measured for an incoherent light source (LED) reflected at the surface of our device. Interference fringes only emerge in the vicinity of the spatial auto-correlation (y=0), thereby providing a direct measurement of our optical resolution (gray area adjusting the variation of the interference contrast on the right panel).}  \\

\newpage

\centerline{\includegraphics[width=\linewidth]{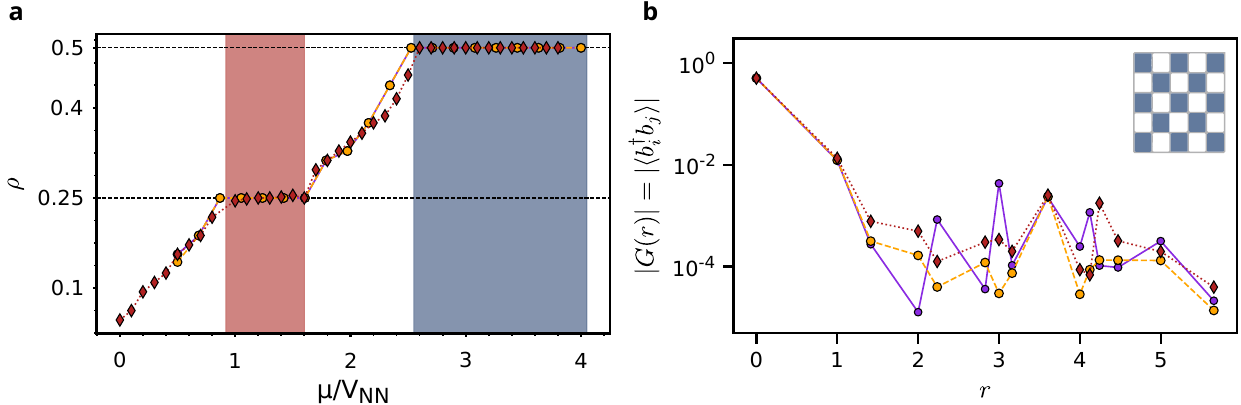}}
\vspace{.1cm}
\textbf{Extended Data Fig.5: Validation for CB and star SS.}  \textbf{a} Particle density $\rho$ as a function of the chemical potential $\mu$ for $J/V_{NN}=0.02$. The shaded rectangles underline the parameter space where solid phases emerge, at $\rho=0.25$ (red) and $\rho=0.5$ (blue). Tensor network results are displayed for all values accessible  to $j_{ij}$ (violet), or negative ones only (yellow), in the latter case compared to QMC (red). \textbf{b} Single particle density matrix $|G(r)|$, as a function of the distance $r=|\textbf{r}_i-\textbf{r}_j|/a$, inside the checkerboard phase for the data displayed in \textbf{a}, at $\mu/V_{NN}=3.5$. \\

\centerline{\includegraphics[width=\linewidth]{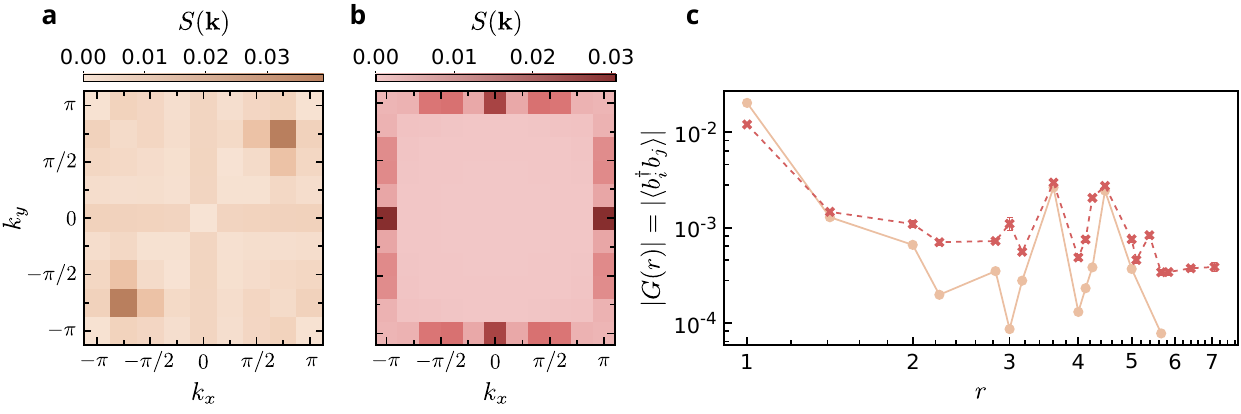}}
\vspace{.1cm}
\textbf{Extended Data Fig.6: Stripe and star SS.} (\textbf{a}-\textbf{b}) Static structure factor $S(\mathbf{k})$, across the first Brillouin zone, for the stripe phase on a $9 \times 9$ lattice with $\mu/V_{NN} = 2.0$ (\textbf{a}), and for the star phase on a $10 \times 10$ lattice with $\mu/V = 1.4$ (\textbf{b}), in both cases for  $J/V_{NN} = 0.02$. The former phase is characterized by a dominant peak at $\mathbf{k} = \pm(2\pi/3, 2\pi/3)$, while the latter is signaled by characteristic peaks near $\mathbf{k} = (0, \pm\pi) \text{ or } (\pm\pi,0)$. For clarity, we set S(\textbf{k}=\textbf{0})=0 in \textbf{a-b}
 (\textbf{c}) Corresponding single-particle density matrix $|G(r)|$, as a function of the distance $r=|\textbf{r}_i-\textbf{r}_j|/a$. 

\centerline{\includegraphics[width=\linewidth]{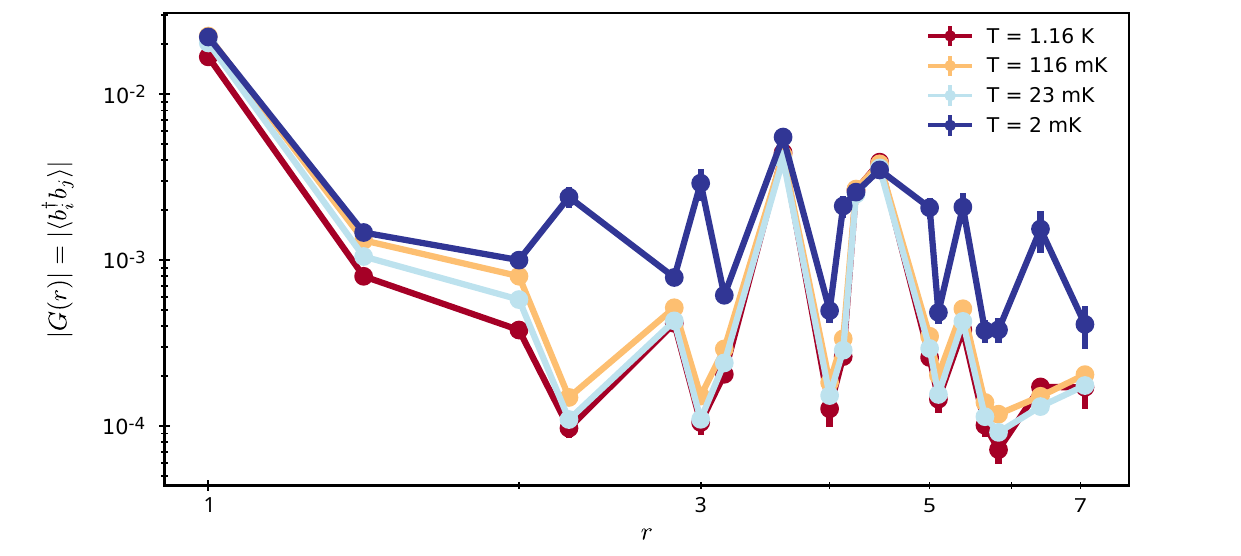}}
\vspace{.1cm}
\fd{\textbf{Extended Data Fig.7: Thermal stability of long-range order.} Decay of the single-particle Green function $|G|$ computed for the CB phase ($\mu/V=4$) and $J/V=0.015$. For temperatures ranging from 2 mK to 1.16 K we verify that the decay of $|G|$ at large distance weakly varies, verifying the robustness of coherent long-range hopping.}\\

\end{document}